\documentclass[12pt,a4paper,final]{revtex4}

\usepackage{sidecap}
\usepackage{ulem}
\usepackage{epsfig}
\usepackage{amsmath,amssymb,amsthm}
\usepackage{graphicx}
\usepackage{bm}
\usepackage{color}

\usepackage{float}

\DeclareMathAlphabet{\mathpzc}{OT1}{pzc}{m}{it}

\begin{document}

\renewcommand{\textfraction}{0.00}


\newcommand{\vAi}{{\cal A}_{i_1\cdots i_n}} 
\newcommand{\vAim}{{\cal A}_{i_1\cdots i_{n-1}}} 
\newcommand{\vAbi}{\bar{\cal A}^{i_1\cdots i_n}}
\newcommand{\vAbim}{\bar{\cal A}^{i_1\cdots i_{n-1}}}
\newcommand{\htS}{\hat{S}} 
\newcommand{\htR}{\hat{R}}
\newcommand{\htB}{\hat{B}} 
\newcommand{\htD}{\hat{D}}
\newcommand{\htV}{\hat{V}} 
\newcommand{\cT}{{\cal T}} 
\newcommand{\cM}{{\cal M}} 
\newcommand{\cMs}{{\cal M}^*}
\newcommand{\vk}{\vec{\mathbf{k}}}
\newcommand{\bk}{\bm{k}}
\newcommand{\kt}{\bm{k}_\perp}
\newcommand{\kp}{k_\perp}
\newcommand{\km}{k_\mathrm{max}}
\newcommand{\vl}{\vec{\mathbf{l}}}
\newcommand{\bl}{\bm{l}}
\newcommand{\bK}{\bm{K}} 
\newcommand{\bb}{\bm{b}} 
\newcommand{\qm}{q_\mathrm{max}}
\newcommand{\vp}{\vec{\mathbf{p}}}
\newcommand{\bp}{\bm{p}} 
\newcommand{\vq}{\vec{\mathbf{q}}}
\newcommand{\bq}{\bm{q}} 
\newcommand{\qt}{\bm{q}_\perp}
\newcommand{\qp}{q_\perp}
\newcommand{\bQ}{\bm{Q}}
\newcommand{\vx}{\vec{\mathbf{x}}}
\newcommand{\bx}{\bm{x}}
\newcommand{\tr}{{{\rm Tr\,}}} 
\newcommand{\bc}{\textcolor{blue}}

\newcommand{\beq}{\begin{equation}}
\newcommand{\eeq}[1]{\label{#1} \end{equation}} 
\newcommand{\ee}{\end{equation}}
\newcommand{\bea}{\begin{eqnarray}} 
\newcommand{\eea}{\end{eqnarray}}
\newcommand{\beqar}{\begin{eqnarray}} 
\newcommand{\eeqar}[1]{\label{#1}\end{eqnarray}}
 
\newcommand{\half}{{\textstyle\frac{1}{2}}} 
\newcommand{\ben}{\begin{enumerate}} 
\newcommand{\een}{\end{enumerate}}
\newcommand{\bit}{\begin{itemize}} 
\newcommand{\eit}{\end{itemize}}
\newcommand{\ec}{\end{center}}
\newcommand{\bra}[1]{\langle {#1}|}
\newcommand{\ket}[1]{|{#1}\rangle}
\newcommand{\norm}[2]{\langle{#1}|{#2}\rangle}
\newcommand{\brac}[3]{\langle{#1}|{#2}|{#3}\rangle} 
\newcommand{\hilb}{{\cal H}} 
\newcommand{\pleft}{\stackrel{\leftarrow}{\partial}}
\newcommand{\pright}{\stackrel{\rightarrow}{\partial}}

\newcommand{\squeezeup}{\vspace{-2.5mm}}


\title{Heavy flavor suppression predictions at 5.1 TeV Pb+Pb collisions at LHC}

\date{\today}
 
\author{Magdalena Djordjevic}
\affiliation{Institute of Physics Belgrade, University of Belgrade, Serbia}

\author{Marko Djordjevic}
\affiliation{Faculty of Biology, University of Belgrade, Serbia}

\begin{abstract}

High momentum hadron suppression is considered to be an excellent probe of jet-medium interactions in QCD matter created in ultra-relativistic heavy ion collisions. We previously showed that our dynamical energy loss formalism can accurately explain suppression measurements at 200 GeV Au+Au collisions at RHIC and 2.76 TeV Pb+Pb collisions at the LHC. With the upcoming LHC measurements at notably higher collision energies, there is a question of what differences, with respect to the current (2.76 TeV) measurements, can be expected. In this paper we concentrate on heavy flavor suppression at the upcoming 5.1 TeV Pb+Pb collisions energy at the LHC. Naively, one would expect a notably ($\sim 30\%$) larger suppression at 5.1 TeV collision energy, due to estimated (significant) energy loss increase when transitioning from 2.76 to 5.1 TeV. Surprisingly, more detailed calculations predict nearly the same suppression results at these two energies. We show that this unexpected result is due to an interplay of the following two effects, which essentially cancel each other: {\it i}) flattening of the initial distributions with increasing collision energies, and {\it ii}) significantly slower than naively expected increase in the energy loss. Therefore, the obtained nearly the same suppression provides a clear (qualitative and quantitative) test of our energy loss formalism.

\end{abstract}

\pacs{12.38.Mh; 24.85.+p; 25.75.-q}
\maketitle 
\section{Introduction} 
High energy heavy flavor suppression~\cite{Bjorken} is considered to be an excellent probe of QCD matter created in ultra-relativistic heavy ion collisions at RHIC and LHC. One of the major goals of these experiments is mapping the QGP properties, which requires comparing available suppression data with the theoretical predictions~\cite{STE, STE1, STE2}. Such comparison tests different theoretical models and provides an insight into the underlying QGP physics. Having this in mind the upcoming 5.1 TeV Pb+Pb measurements at LHC (expected at the end of 2015) - and their comparison with theoretical predictions - will provide an additional important insight in the jet-medium interactions in QGP created in such collisions. With this motivation in mind, the goal of this paper is providing the heavy flavor suppression predictions, and physical interpretation behind the obtained results, for the upcoming high-luminosity experimental data at 5.1 TeV Pb+Pb collisions at LHC. In particular, we aim assessing the differences in the predicted suppression with respect to the already available 2.76 TeV measurements at LHC and compare the results of state-of-the-art calculations with simple expectations/estimates.

To generate the theoretical predictions  we will use our recently developed dynamical energy loss formalism, which  includes: \textit{i)} dynamical scattering centers, \textit{ii)} QCD medium of a finite size~\cite{DynEL, DynEL1}, \textit{iii)} both radiative~\cite{DynEL, DynEL1} and collisional~\cite{MD_Coll} energy losses, \textit{iv)} finite magnetic mass effects~\cite{MagM} and \textit{v)} running coupling~\cite{RunnC}. This energy loss formalism is based on the pQCD calculations in finite size optically thin dynamical QCD medium, and has been incorporated into a numerical procedure~\cite{RunnC} that allows generating state-of-the art suppression predictions. The model has shown to be successful in explaining a wide range of angular averaged observables~\cite{HFLHC,RunnC,CRHIC,NCLHC} at both RHIC and LHC. Since the angular averaged $R_{AA}$s are largely insensitive to the medium evolution, angular averaged $R_{AA}$ can be considered an excellent probe for jet-medium interactions~\cite{Thorsten,Molnar,Footnote1}; consequently, the suppression predictions at 5.1 TeV Pb+Pb collisions at LHC, and their comparison with the measurements, will allow further testing of our energy loss formalism.

\section{Overview of the computational framework}

For generating the suppression predictions, we use the computational procedure from~\cite{RunnC}. The main features are briefly summarized below, while the full account of the procedure is provided in~\cite{RunnC}.  

The  quenched spectra of heavy flavor observables are calculated according to the generic pQCD convolution:
\begin{eqnarray}
\frac{E_f d^3\sigma}{dp_f^3} = \frac{E_i d^3\sigma(Q)}{dp^3_i}
 \otimes
{P(E_i \rightarrow E_f )}
\otimes D(Q \to H_Q) \otimes f(H_Q \to e, J/\psi). \; 
\label{schem} \end{eqnarray}

In the equation above subscripts "i"  and "f" correspond, respectively, to "initial" and "final", and $Q$ denotes heavy quarks. $E_i d^3\sigma(Q)/dp_i^3$ denotes the initial heavy quark spectrum, which is computed at next to leading order according to~\cite{Cacciari:2012,FONLL}.  $P(E_i \rightarrow E_f )$ is the energy loss probability; this probability includes both radiative and collisional energy loss in a 
finite size dynamical QCD medium, multi-gluon~\cite{GLV_suppress} and path-length fluctuations~\cite{WHDG} and running coupling~\cite{RunnC}. $D(Q \to H_Q)$ is the fragmentation function of heavy quark $Q$ to hadron $H_Q$, where for D and B mesons we use, BCFY~\cite{BCFY} and KLP~\cite{KLP} fragmentation functions, respectively. Finally, decay of B mesons to experimentally measured non-prompt $J/\psi$ is represented by $f(H_Q \to J/\psi)$ and obtained according to~\cite{Cacciari:2012}.  
\medskip

The expression for the radiative energy loss in a finite size dynamical QCD medium is extracted from Eq. (10) in~\cite{MagM}, while the collisional energy loss is extracted from Eq. (14) in~\cite{MD_Coll}. Path length distributions are taken from~\cite{Dainese}.

The angular averaged $R_{AA}$ is a clear jet-medium interaction probe, i.e. it is not sensitive on the details of the medium evolution~\cite{Molnar,Thorsten}, so we model the medium by assuming constant average temperature of QGP. To determine the average temperatures at 0-10\% most central collisions, we start from $T{\,=\,}304$\,MeV (the effective temperature extracted by ALICE~\cite{LHC_T} for 0-40\% centrality), and use the procedure outlined in~\cite{NCLHC} (based on gluon rapidity density) to determine the temperatures at central collisions at 2.76 and 5.1 TeV Pb+Pb collisions; for 2.76 TeV 0-10\% centrality, this leads to the average temperature of 313 MeV.  To determine the temperature at 5.1 TeV, note that it is expected that the gluon rapidity density will be 25\% higher at 5.1 TeV than at 2.76 TeV in Pb+Pb collisions at LHC~\cite{LHC_Mult}. Since the temperature is proportional to the gluon rapidity density, i.e. $T \sim (dN_g/dy)^{1/3}$, this leads to $\sim 7\%$ higher temperature at 5.1 TeV compared to 2.76 TeV at the LHC, i.e. 335 MeV for 0-10\% central 5.1 TeV Pb+Pb collisions at LHC. Note that, in our energy loss calculations, this is the only parameter that differs between the two systems; i.e. all the other parameters that enter in the calculations (stated in the next paragraph) are the same for the two systems, and correspond to the standard literature values (i.e. no parameters are determined through fitting the data).

The following parameters are used in the numerical calculations: QGP with effective light quark flavors $n_f{\,=\,}3$ and perturbative QCD scale of $\Lambda_{QCD}=0.2$~GeV. The Debye mass is taken to be $\mu_E \approx 0.9$ ($\mu_E \approx 0.97$) GeV for 2.76 (5.1) TeV collision energy, and is obtained by self-consistently solving  Eq.~(7) in~\cite{Peshier}. The value for magnetic to electric mass ratio $\mu_M/\mu_E$ is extracted from non-perturbative calculations~\cite{xb1,xb2,xb3,xb4} $0.4 < \mu_M/\mu_E < 0.6$; the gluon mass is  $m_g=\mu_E/\sqrt{2}$~\cite{DG_TM}, while the 
charm  and the bottom mass are, respectively, $M{\,=\,}1.2$\,GeV  and $M{\,=\,}4.75$\,GeV. Path-length distribution, parton production, fragmentation functions and decays, which are used in the numerical calculations, are specified above.

\section{Results and discussion}

To get an insight of what results we expect at 5.1 TeV collisions at the LHC, we will first provide a simple analytic estimate for heavy flavor suppression at this collision energy. For that purpose, note that radiative energy loss is widely considered to be a dominant energy loss mechanism in QGP, so we will use only the radiative contribution for the estimate. Since it is also widely assumed that radiative energy loss is proportional to $T^3$ (see e.g.~\cite{Betz}), one can estimate that the energy loss at 5.1 TeV should be $ \sim 25\%$ higher than at 2.76 TeV. Based on this, and if we assume that initial distributions can be approximated by power low distributions, i.e. $d\sigma/dp_\perp^2 \sim 1/p_\perp^n$, we can make an estimate on how much larger/smaller suppression one would expect at 5.1 TeV compared to the already observed results at 2.76 TeV. 

It was previously shown that, for radiative energy loss and power low initial distributions, suppression can be roughly estimated by using the following simple formula~\cite{GLV_suppress,Footnote2}:
\begin{eqnarray}
(1-\frac{1}{2} \frac{\Delta E}{E})^{(n-2)} &\approx& (1-\frac{n-2}{2} \frac{\Delta E}{E}),
\label{suppSimpolified}
\end{eqnarray}
where $\Delta E/E$ is the fractional energy loss. If we assume that, at 2.76 TeV, typical fractional energy loss for charm is $\sim 30\%$ and for bottom $\sim 15\%$, and that charm and bottom distributions do not notably change between these two collision energies, with $n \sim 6.5$ ($n \sim 6$) for charm (bottom), the above estimate will straightforwardly lead to the expectation of $\sim 30 \%$ ($\sim 10 \%$) larger suppression for charm (bottom) at 5.1 TeV compared to 2.76 TeV Pb+Pb collisions at the LHC.  

\begin{figure*}
\epsfig{file=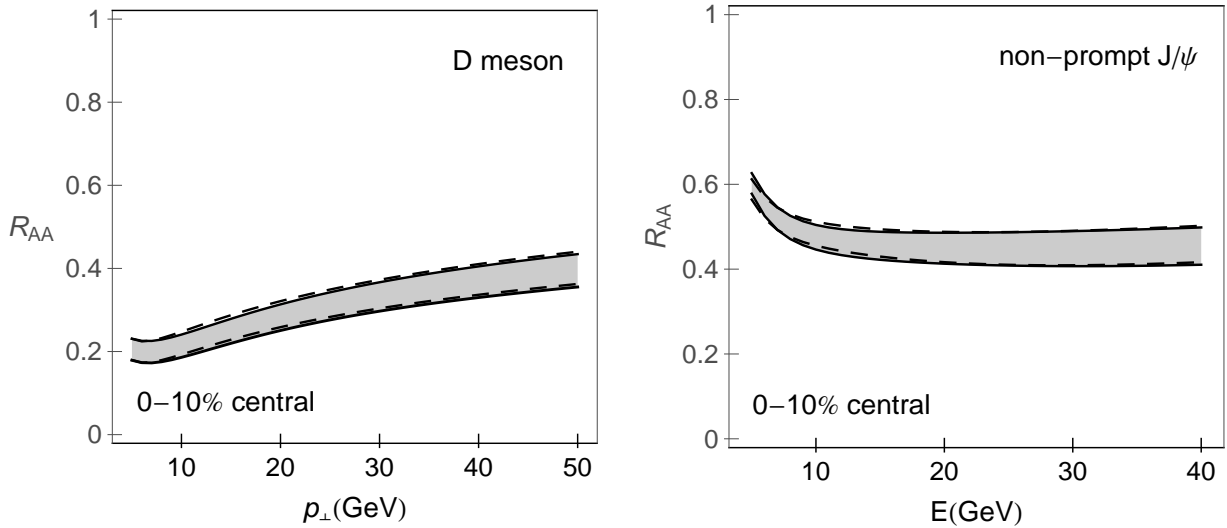,scale=0.65}
\vspace*{-0.5cm}
\caption{\textbf {Comparison of $R_{AA}$ predictions for heavy flavor at 2.76 and 5.1 TeV.} D mesons and non-prompt $J/\psi$ suppression predictions, as a function of transverse momentum, are shown on the left and the right panel, respectively. Full (dashed) curves correspond to $R_{AA}$ predictions at 5.1 TeV (2.76 TeV) Pb+Pb collisions at the LHC. On each panel, the gray bands correspond to the finite magnetic mass case (i.e.  $0.4 < \mu_M/\mu_E < 0.6$~\cite{xb1,xb2,xb3,xb4}), where the lower and the upper boundary correspond, respectively, to $\mu_M/\mu_E=0.4$ and $\mu_M/\mu_E =0.6$. }
\label{SuppFig1}
\end{figure*}

Contrary to these expectations, Figure~\ref{SuppFig1} shows that our suppression calculations - obtained from the energy loss formalism outlined in the previous section - provide substantially different predictions. From this figure, we actually do not observe any suppression increase between 2.76 to 5.1 TeV collisions at the LHC. That is, we obtain the same suppression patterns for both charm and bottom probes (D mesons and non-prompt $J/\psi$) at these two collision energies. This then leads to the question of, why the increase in the collision energy by almost a factor of 2, leads to the same predicted suppression patterns between the two collisional energies, despite the estimated significant (i.e. $\sim 30\%$ for charm, see above) increase in the suppression? 

\begin{figure*}
\epsfig{file=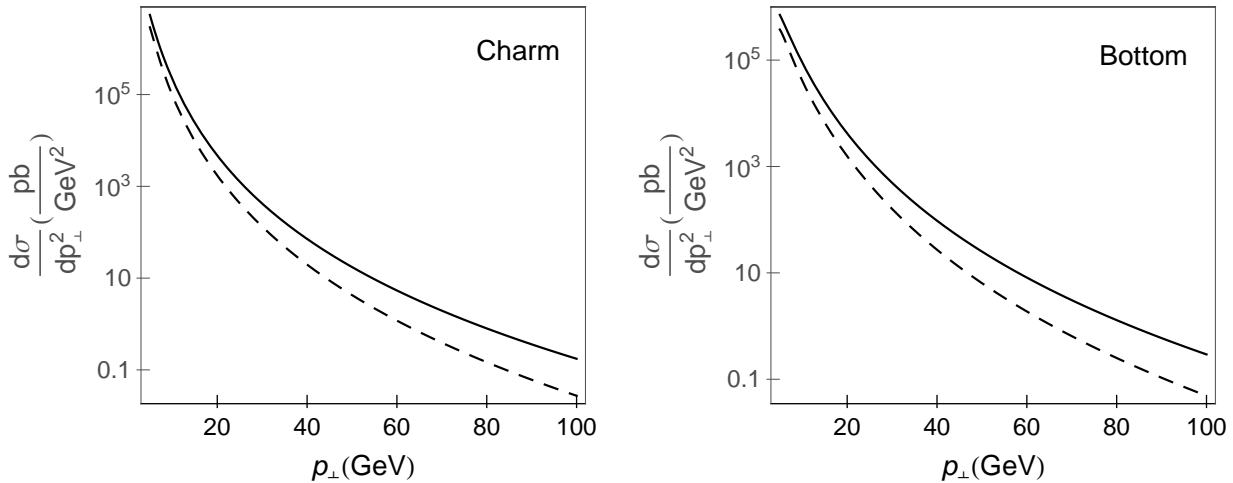,scale=0.65}
\vspace*{-0.5cm}
\caption{\textbf {Comparison of momentum distributions for charm and bottom at 2.76 and 5.1 TeV.} Charm and bottom momentum distributions, as a function of transverse momentum, are shown on the left and the right panel, respectively. On each panel full (dashed) curve corresponds to the momentum distribution at 5.1 TeV (2.76 TeV) Pb+Pb collisions at the LHC. }
\label{ptdist}
\end{figure*}

\begin{figure*}
\epsfig{file=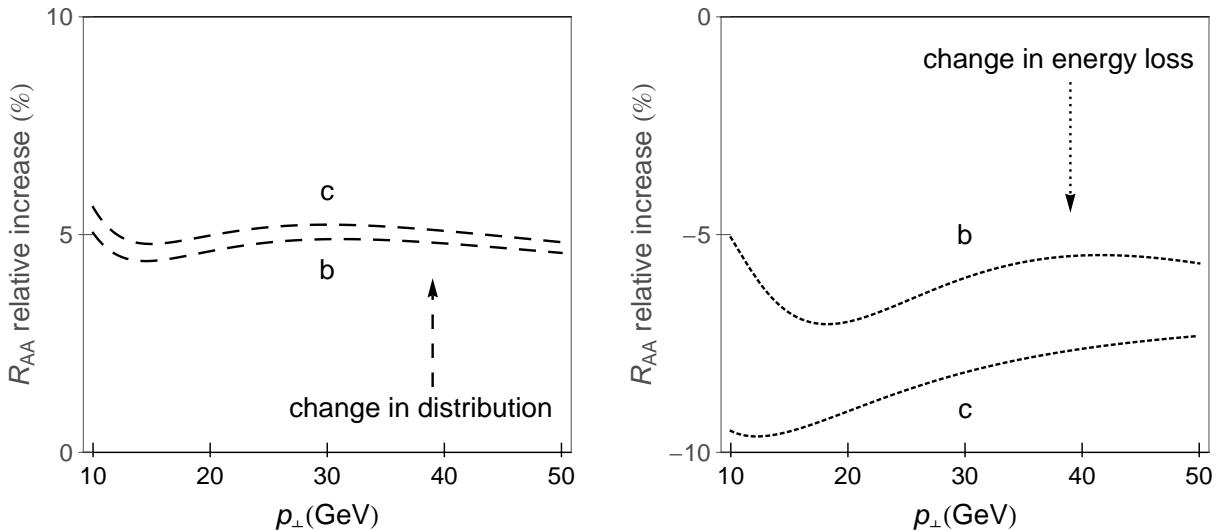,scale=0.65}
\vspace*{-0.5cm}
\caption{\textbf {Relative increase in $R_{AA}$ between 2.76 and 5.1 TeV.} The left panel shows momentum dependence of the relative increase in $R_{AA}$ between 2.76 and 5.1 TeV Pb+Pb collisions at the LHC due to differences in the  distributions; the energy loss is kept fixed and calculated for 2.76 TeV case. The right panel shows momentum dependence of the relative increase in $R_{AA}$ between 2.76 and 5.1 TeV collisions at the LHC due to differences in the energy loss; the momentum distribution is kept fixed and calculated for 2.76 TeV case. On each panel, curves that correspond to charm and bottom are marked by c and b, respectively, and the magnetic mass is fixed to  $\mu_M/\mu_E=0.4$. }
\label{DecreaseIncrease}
\end{figure*}

To address this question, in Fig.~\ref{ptdist} we  first compare charm and bottom initial distributions between these two collision energies. From this figure, we see that the distributions at 5.1 TeV are slightly flatter than at 2.76 TeV, for both charm and bottom, which will have the tendency to somewhat lower the suppression at 5.1 TeV compared to 2.76 TeV. Note that only the shape of the distributions contributes to the suppression predictions, and from Fig.~\ref{ptdist}, one can observe that the differences in the shape of the distributions  are not large. Still, this difference in the distributions has a notable (though again not large, i.e. $\sim 5 \%$) effect on the suppression predictions, as can be seen in the left panel of Fig.~\ref{DecreaseIncrease}; therefore, it should be taken into account in the suppression calculations. 

However, what we further see from the right panel of Fig.~\ref{DecreaseIncrease} is that the effect on the suppression coming from the energy loss increase between 2.76  and 5.1 TeV (due to the increase in average temperature) is also notable but not large, i.e. it corresponds to $5$ and $10 \%$. That is, the energy loss effect on the suppression has about the same magnitude, but an opposite direction, compared to the effect of different distributions between these two collision energies. The first question that we want to address is why the effect of the energy loss increase on jet suppression is not larger between these two collision energies, at least not for charm quark. That is, based on the common $T^3$ assumption, we have estimated that the energy loss increase should be on the order of $25\%$, which should, therefore, have a more prominent (estimated ~30\%) effect on the suppression. 

Regarding the $T^3$ estimate for the radiative energy loss, note that, while widely used, this estimate does not have to be justified. That is, from Eq. (10) in \cite{MagM}, which shows the radiative energy loss expression in a finite size dynamical QCD medium, it can be straightforwardly observed that the expression nontrivially depends on T. That is, while one can recover  a part with explicit dependence on $T^3$ in this expression, the rest of the expression also depends on T, where this extra term considerably  modifies the temperature dependence. Additionally, the collisional energy loss effect, while smaller compared to the radiative, is still important, and this effect also has to be taken into account in the suppression calculations. Note that, for the collisional energy loss, it is commonly assumed that it has a quadratic ($T^2$) dependence on the temperature. However, similarly to the above discussion for the radiative energy loss, Eq. (14) from \cite{MD_Coll} shows  a nontrivial temperature dependence, so we will below also test whether this simple ($T^2$) assumption is justified. 

\begin{figure*}
\epsfig{file=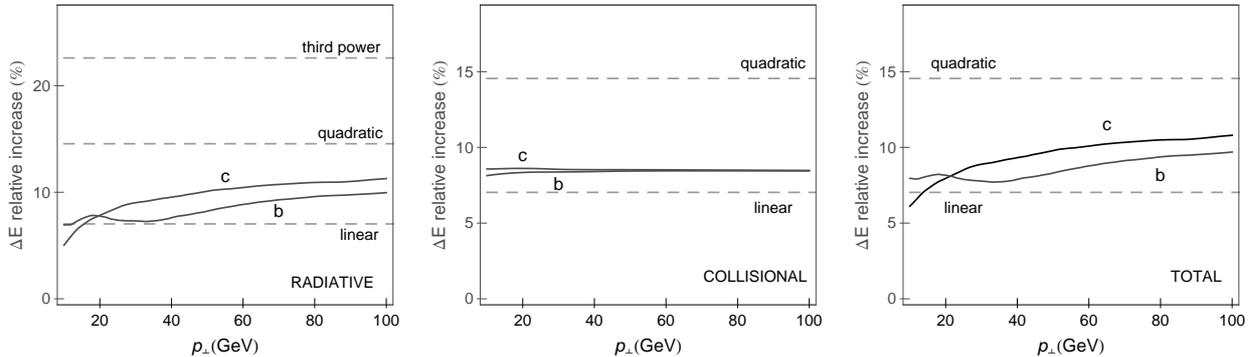,scale=0.42}
\vspace*{-0.5cm}
\caption{\textbf {Relative energy loss increase between 2.76 and 5.1 TeV.} All the  panels show the momentum dependence of the relative energy loss increase between 5.1 and 2.76 TeV Pb+Pb collisions at the LHC. The left, the central and the right panel correspond, respectively, to the radiative, collisional and total energy loss case. On each panel curves that correspond to charm (bottom) are marked by letter c (b) and the  magnetic mass is fixed to  $\mu_M/\mu_E=0.4$. Dashed gray horizontal lines represent the energy loss increase, if it would have linear, quadratic or cubic temperature dependence.}
\label{ELossRatioDiffTemp}
\end{figure*}

With these goals, in Fig.~\ref{ELossRatioDiffTemp}, we plot the relative heavy flavor energy loss increase between 2.76 and 5.1 collision energies at the LHC. Figures also contain dashed horizontal lines, which represent what would be the energy loss increase, if it would indeed have $T^2$ or $T^3$ dependence. For radiative energy loss, we see that, contrary to the common expectations, energy loss increase is far from $T^3$ dependence; i.e. it is between linear (for low jet energy regions) and quadratic (which can be reached for asymptotically high jet energies). Consequently, for the high momentum heavy flavor hadrons that will be studied at these two collision energies at the LHC, the expected energy loss increase is notably smaller than quadratic, i.e. it is in the region between  $5-10\%$ (note that the average temperature increase between these two collision energies is $\sim 7\%$). For the collisional energy loss, we also see that energy loss increase is far from quadratic, i.e. the increase of $\sim 8.5 \%$ is constant with momentum and it has slightly larger than linear dependence on temperature. Consequently, contrary to the common expectation, the total energy loss has also a modest temperature dependence, which is close to linear, i.e. between 6 and 10\% depending on the jet momentum. This modest energy loss increase between these two collisional energies consequently leads to a modest increase in the suppression which we observe in the right panel of Fig.~\ref{DecreaseIncrease}.
  
\begin{figure*}
\epsfig{file=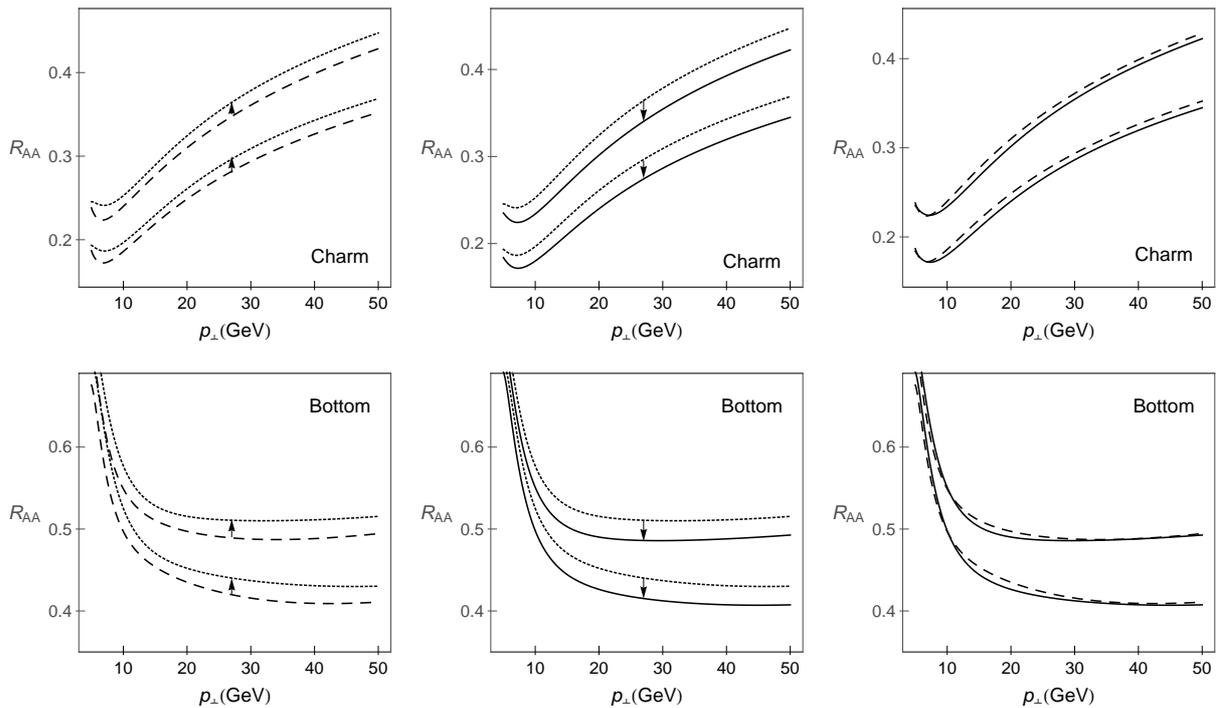,scale=0.45}
\vspace*{-0.5cm}
\caption{\textbf {Analysis of the heavy flavor suppression between 2.76 and 5.1 TeV.} The full curves correspond to $R_{AA}$s with both the energy losses and the distributions calculated at 5.1 TeV collision energy. The dashed curves correspond to $R_{AA}$s with both the energy losses and the distributions calculated at 2.76 TeV collision energy. The dotted curves correspond to $R_{AA}$s where the energy losses are calculated at 2.76 TeV collision energy, while the distributions are calculated at 5.1 TeV collision energy. Upper (lower) panels correspond to the charm (bottom) quark. Left panels show how the flatter distributions at 5.1 TeV lower the heavy flavor suppression compared to the 2.76 TeV case. Central panels show how increase in the energy loss at 5.1 TeV increase the suppression compared to the 2.76 TeV case. Right panels show how the above two effects cancel, so as to reproduce almost the same suppression at 2.76 and 5.1 TeV Pb+Pb collision energy. On each panel, lower (upper) set of curves correspond to the magnetic to electric mass ratio of $\mu_M/\mu_E=0.4$ ( $\mu_M/\mu_E=0.6$). }
\label{Suppression}
\end{figure*}

Finally, in Fig.~\ref{Suppression}, we study the combined effect of the differences in the distributions and the energy loss on jet suppression. On the two left panels, we see the effect of the difference in the distributions on the jet suppression, while the energy loss is kept fixed. On the two central panels, we keep the same distribution, but change the energy loss, while in the two right panels both the distributions and the energy loss are changed between the two collision energies. From the panels, we see that, while the change in the distribution has the tendency to reduce the suppression, the energy loss increase increases the suppression for about the same amount, so that the resultant suppression at 5.1 TeV collision energy is almost the same as at 2.76 TeV. 

The above obtained numerical result can also be directly estimated from Eq.~\ref{suppSimpolified}. For this purpose, we will take that the energy loss between 2.76 and 5.1 TeV collision energy increases by factor $\eta$, where from Fig.~\ref{ELossRatioDiffTemp}, we see that $\eta \approx 10\%$ for both charm and bottom. Additionally, we will take that the power factor in the initial parton distributions decrease by $\delta$; by fitting the power low to the ratio of the momentum distributions in Fig.~\ref{ptdist}, we obtain $\delta \approx 0.4$. By applying these factors into Eq.~\ref{suppSimpolified}, one can straightforwardly obtain
\begin{eqnarray}
R_{AA} (5.1 \, {\rm TeV}) \approx R_{AA} (2.76\, {\rm TeV})+\frac{1}{2}\frac{\Delta E}{E}(\delta-\eta (n-2)),
\label{suppComparison}
\end{eqnarray} 
where for $\delta$ and $\eta$ estimated above the second additive in the above becomes close to zero. Consequently, this estimate also recovers the conclusion of the same heavy flavor suppression at 2.76 and 5.1 TeV Pb+Pb collision energies at the LHC.

\section{Conclusion}

In this paper, we provided heavy flavor suppression predictions for the upcoming 5.1 TeV Pb+Pb collisions at the LHC. Based on our energy loss formalism, we predict the same heavy flavor suppression patterns for 2.76 and 5.1 TeV collision energies. This result is surprising since, based on the commonly used assumption, a notable increase of the suppression is expected at the higher collision energy. We showed that the same suppression is a consequence of the interplay between the following two effects: {\it i}) a decrease in the suppression due to flattening of the initial momentum distributions, and {\it ii}) an increase in the suppression - though more moderate than expected - due to higher energy loss. Consequently, this unexpected, but simple, suppression prediction provides a direct (both quantitative and qualitative) test of our understanding of the medium interactions in QCD medium created in these collisions.

\bigskip

{\em Acknowledgments:} 
This work is supported by Marie Curie International Reintegration Grant 
within the $7^{th}$ European Community Framework Programme 
PIRG08-GA-2010-276913 and by the Ministry of Science and Technological 
Development of the Republic of Serbia, under projects No. ON171004 and ON173052.


\begin{references}
\bibitem{Footnote1} This is in distinction to high-momentum elliptic flow, which is sensitive to the medium evolution, and therefore is a main probe for the properties of the medium created in ultra-relativistic heavy ion collisions.
\bibitem{Footnote2} This formula should not be used for reliable predictions - we here use it only for the purpose of analytical estimate.
\end{references}
\end{document}